\documentclass[10pt,conference]{IEEEtran}

\usepackage{amsmath,times,amssymb,epsfig,nicefrac,color}
\usepackage{amsfonts,graphics,subfigure,epsfig,graphics, mathabx}
\usepackage[mathcal]{euscript} % redefine \mathcal to be \EuScript;
\usepackage{mathrsfs}
\usepackage{cite,calc}
\usepackage{euler}
\newcommand\nc\newcommand
\nc\bfa{{\boldsymbol a}}\nc\bfA{{\boldsymbol A}}\nc\cA{{\mathcal A}}
\nc\bfb{{\boldsymbol b}}\nc\bfB{{\boldsymbol B}}\nc\cB{{\mathcal B}}
\nc\bfc{{\boldsymbol c}}\nc\bfC{{\boldsymbol C}}\nc\cC{{\mathcal C}}
\nc\sC{{\mathscr C}}
\nc\bfd{{\boldsymbol d}}\nc\bfD{{\boldsymbol D}}\nc\cD{{\mathcal D}}
\nc\bfe{{\boldsymbol e}}\nc\bfE{{\boldsymbol E}}\nc\cE{{\mathcal E}}
\nc\bff{{\boldsymbol f}}\nc\bfF{{\boldsymbol F}}\nc\cF{{\mathcal F}}
\nc\bfg{{\boldsymbol g}}\nc\bfG{{\boldsymbol G}}\nc\cG{{\mathcal G}}
\nc\bfh{{\boldsymbol h}}\nc\bfH{{\boldsymbol H}}\nc\cH{{\mathcal H}}
\nc\bfi{{\boldsymbol i}}\nc\bfI{{\boldsymbol I}}\nc\cI{{\mathcal I}}
\nc\bfj{{\boldsymbol j}}\nc\bfJ{{\boldsymbol J}}\nc\cJ{{\mathcal J}}
\nc\bfk{{\boldsymbol k}}\nc\bfK{{\boldsymbol K}}\nc\cK{{\mathcal K}}
\nc\bfl{{\boldsymbol l}}\nc\bfL{{\boldsymbol L}}\nc\cL{{\mathcal L}}
\nc\bfm{{\boldsymbol m}}\nc\bfM{{\boldsymbol M}}\nc\sM{{\mathscr M}}
\nc\bfn{{\boldsymbol n}}\nc\bfN{{\boldsymbol N}}\nc\cN{{\mathcal N}}
\nc\bfo{{\boldsymbol o}}\nc\bfO{{\boldsymbol O}}\nc\cO{{\mathcal O}}
\nc\bfp{{\boldsymbol p}}\nc\bfP{{\boldsymbol P}}\nc\cP{{\mathcal P}}
\nc\bfq{{\boldsymbol q}}\nc\bfQ{{\boldsymbol Q}}\nc\cQ{{\mathcal Q}}
\nc\bfr{{\boldsymbol r}}\nc\bfR{{\boldsymbol R}}\nc\cR{{\mathcal R}}
\nc\bfs{{\boldsymbol s}}\nc\bfS{{\boldsymbol S}}\nc\cS{{\mathcal S}}
\nc\bft{{\boldsymbol t}}\nc\bfT{{\boldsymbol T}}\nc\cT{{\mathcal T}}
\nc\bfu{{\boldsymbol u}}\nc\bfU{{\boldsymbol U}}\nc\cU{{\mathcal U}}
\nc\bfv{{\boldsymbol v}}\nc\bfV{{\boldsymbol V}}\nc\cV{{\mathcal V}}
\nc\bfw{{\boldsymbol w}}\nc\bfW{{\boldsymbol W}}\nc\cW{{\mathcal W}}
\nc\bfx{{\boldsymbol x}}\nc\bfX{{\boldsymbol X}}\nc\cX{{\mathcal X}}
\nc\bfy{{\boldsymbol y}}\nc\bfY{{\boldsymbol Y}}\nc\cY{{\mathcal Y}}
\nc\bfz{{\boldsymbol z}}\nc\bfZ{{\boldsymbol Z}}\nc\cZ{{\mathcal Z}}

\DeclareMathOperator{\wt}{wt}

\newcommand{\h}{h_\mathrm{B}}
\newcommand{\avg}{{\mathbb E}}
\newcommand{\dist}{d_\mathrm{H}}

\newtheorem{theorem}{Theorem}
\newtheorem{definition}{Definition}
\newtheorem{lemma}[theorem]{Lemma}

\newcommand\ff{{\mathbb F}}

\begin{document}
\title{On the Capacity of Memoryless  Adversary}
\author{\IEEEauthorblockN{Arya Mazumdar}
\IEEEauthorblockA{Department of ECE\\
University of Minnesota--
Twin Cities \\Minneapolis, MN  55455\\ email: \texttt{arya@umn.edu}}
}
%\thanks{This work was supported in part by the some grants }

\maketitle
% \thanks{This work was supported in part by NSF grant 1318093 and a grant from University of Minnesota.}

{\renewcommand{\thefootnote}{}\footnotetext{

\vspace{.02in}This work was supported in part by NSF grant 1318093 and a grant from University of Minnesota.
}}
\renewcommand{\thefootnote}{\arabic{footnote}}
\setcounter{footnote}{0}\maketitle

%\let\thefootnote\relax\footnotetext{This work was supported in part by NSF grant 1318093 and a grant from University of Minnesota.}
%\thispagestyle{empty}
%\begin{abstract}

%  \end{abstract}
%  \newpage
 % \pagestyle{plain}
 %\pagenumbering{arabic} 
 
 \allowdisplaybreaks
% 1. WHAT CAN BE THE MAXIMUM SIZE OF A CODE WITH HEAVY TAILED DISTANCE DISTRIBUTION? NO CONSTRAINS ON THE DISTANCE. USE LINEAR PROGRAM BOUND.
 
 \begin{abstract}
 In this paper, we study a model of communication under adversarial noise. In this model, the
 adversary 
 makes online decisions on whether to corrupt a transmitted bit based on only
 the value of that bit.  Like the usual binary symmetric channel of information theory 
 or the fully adversarial channel of
 combinatorial coding theory, the adversary can, with high probability, introduce at most a
 given fraction of error.
 %knows the codebook and sees only the transmission of 
 
 It is shown that, the capacity (maximum rate of reliable information transfer) of such
 memoryless adversary is strictly below that of the binary symmetric channel. 
 We give new upper bound on the capacity of such channel -- the tightness of this upper bound
  remains an open question. The main component of our proof is the careful examination of
 error-correcting properties of a code with skewed distance distribution.
 %The tightness of our
 %upper bound on the capacity of memoryless adversary is open for study and we make
 %several steps towards the solution of this problem.
 \end{abstract}

 \section{Introduction}
 Consider the usual definitions of discrete channels in information theory. It is assumed that,
 transmissions of symbols from a discrete alphabet take place and a fraction of the transmissions may result in erroneous reception. 
 The sender is allowed to ``encode'' information in to an array of symbols, called a codeword. The collection of
 all possible codewords is called a ``code'' (or ``codebook''). Without much loss of  generality, we can assume that all transmitted codewords
 are equally likely, in which case  the log-size of a code signify the amount of information that can be transmitted with the code.
 In a completely adversarial channel, the adversary is allowed to 
see the transmitted set of symbols (codeword) completely and then  decides which of the transmitted symbols are to
be corrupted (it is allowed to corrupt a given fraction of all symbols).

 Recently, in a series of papers \cite{langberg2009binary,dey2013upper,haviv2011beating},  the study of
 online or causal adversarial channels is  initiated, in particular, for binary-input channels.  
 Let us start by giving an informal definition of a causal adversarial channel. 
In the causal adversarial model, an adversary is allowed to see
 the transmitted codeword only causally (i.e., at any instance it sees only the past transmitted symbols), and decides whether to corrupt the current transmitted symbol.
An upper bound on the capacity (maximum rate of reliable information transfer) 
of such channel is presented in \cite{dey2013upper}. One of the most
interesting observation is that, such channels are limited by the ``Plotkin bound,'' of coding theory: whenever
the fraction of error introduced by the adversary surpasses $\frac14$, the capacity is zero (assuming binary input). On the
other hand, by ``random coding'' method, a lower bound is established in \cite{haviv2011beating}.
This 	lower bound beats the famous Gilbert-Varshamov bound, the best available lower bound for a 
completely adversarial channel. 
 
We below describe an adversarial channel model that
is weaker (in terms of adversary limitations) than the above causal channel. 
In particular, the adversary is not even allowed to see the past 
transmitted symbols, but decides whether to corrupt a symbol based on only
the current transmission.
Our initial aim is to see
%how weak the adversary can be with
whether the channel capacity is still dictated by the Plotkin bound. 
 
 \subsection{A memoryless (truly online) adversary}
 In this work we consider the code to be {\em deterministic}, in a sense that is described below. Also,
 we assume that the input alphabet to be binary ($\{0,1\}$).
%The results extend for stochastic codes as well. 
A code $\cC$ is simply a subset of $\ff_2^n$.
The size of the code denotes the number of messages encodable with this code; and therefore
the amount of information encodable is $\log |\cC|$. In here and subsequently, all logarithms are base-2, unless otherwise mentioned.
The {\em rate} of the code is $\frac{\log |\cC|}{n}$.

Given the code, the adversarial channel consists of $n$ (possibly random) functions $f_{\cC}^i:\ff_2 \to \ff_2, i=1,\dots,n.$ 
Suppose a randomly and uniformly chosen codeword $$\bfx \equiv(x_1,x_2,\ldots,x_n) \in \cC$$ is transmitted. At the $i$th time instant, the
adversary will produce $e_i = f_{\cC}^i(x_i)$, taking  only the  current transmitted symbol $x_i$ as argument (and of course,
taking into account the code $\cC$, which is known to the adversary). For, $i = 1,\dots,n$, $e_i$ is the indicator of an error at the
$i$th position. I.e., the channel produces $y_i =x_i+e_i$, at the $i$th time-instance, where the addition is of course over $\ff_2$.
\begin{definition}
 The adversary is called {\em weakly-$p$-limited}, $0\le p\le 1$, if the
expected (with respect to the randomness in $f_{\cC}^i$s and  $\bfx$) Hamming weight of the error-vector $\bfe = (e_1,e_2, \dots, e_n) = 
( f_{\cC}^1(x_i),\dots, f_{\cC}^n(x_n)) \equiv  f_{\cC}(\bfx)$ is 
\begin{equation}
\avg\wt(\bfe) \le pn.
\end{equation}
A more restrictive adversary ({\em strongly-$p$-limited}) must have,
\begin{equation}\label{eq:pn}
\Pr(\wt(\bfe)/n <  p+\epsilon )  = 1- o(1), \forall \epsilon >0.
\end{equation}
\end{definition}
A code is associated with a (possibly randomized) decoder $\phi: \ff_2^n \to \cC$. For a given pair of transmitted codeword and error vector, $ \bfx \in \cC, \bfe\in \ff_2^n$, the  decoder makes
an error if, $\phi(\bfx +\bfe) \ne \bfx$.
Given $\cC$ and $p$, define ${\rm Adv}_w(\cC,p)$ to be the collection of all weakly-$p$-limited adversary strategies. That is,
$f_{\cC}\equiv \{f_{\cC}^i:\ff_2 \to \ff_2, i=1,\dots,n\} \in {\rm Adv}_w(\cC,p)$ if and only if, $\avg\wt(f_{\cC}(\bfx))\le pn.$
Similarly, we can name the collection of all strongly-$p$-limited adversary strategies as ${\rm Adv}_s(\cC,p)$.

Our results, as in the case of causal adversarial channels of \cite{langberg2009binary}, holds for the case
of {\em average probability of error} \footnote{It is relatively easy to see that the worst-case probability of error
does not lead to anything different than the completely adversarial channel. For the same reason linear
codes do not lead to any improvement for these channels over completely adversarial
channel. We refer to \cite{dey2013upper} for further discussion. In general, the notion of average vs. worst-case 
error probability leading to different capacities for {\em arbitrarily varying channels} 
is well-known (for example, see \cite{ahlswede1970capacity} or \cite{lapidoth1998reliable}).}. 

%Suppose, $\bfx \in \cC$ is a randomly and uniformly chosen codeword
%from $\cC$. 
The average probability of error   
is defined to be,
$$
P^w_{\cC}(p) =  \max_{f_{\cC}\in {\rm Adv}_w(\cC,p) } \frac{1}{|\cC|} \sum_{\bfx \in \cC}   \Pr(\phi(\bfx+f_{\cC}(\bfx)) \ne \bfx),
$$
and,
$$
P^s_{\cC}(p) =  \max_{f_{\cC}\in {\rm Adv}_s(\cC,p) } \frac{1}{|\cC|} \sum_{\bfx \in \cC}   \Pr(\phi(\bfx+f_{\cC}(\bfx)) \ne \bfx).
$$
 
The maximum possible size of  ``good'' codes are:
\begin{equation}
M^w_\epsilon(n,p) \equiv \max_{\cC\subseteq\ff_2^n:  P^w_{\cC}(p)\le \epsilon} |\cC|,
\end{equation}
and,
\begin{equation}
M^s_\epsilon(n,p) \equiv \max_{\cC\subseteq\ff_2^n:  P^s_{\cC}(p)\le \epsilon} |\cC|. 
\end{equation}
Now, define the {\em capacities} to be,
\begin{equation}
C_w(p) \equiv \inf_{\epsilon>0} \limsup_{n\to \infty} \frac{\log M^w_\epsilon(n,p)}{n},
\end{equation}
\begin{equation}
C_s(p) \equiv \inf_{\epsilon>0} \limsup_{n\to \infty} \frac{\log M^s_\epsilon(n,p)}{n}.
\end{equation}
%We need to distinguish this from the capacity of the strongly-limited adversary, which can 
%be defined in the same manner and can be denoted as $C_s(p)$.

It is evident that,
\begin{equation}
C_w(p)  \le C_s(p) \le 1-\h(p),
\end{equation}
where $\h(x) = -x \log x -(1-x)\log(1-x)$ is the {\em binary entropy function}.

%\begin{lemma}
%$$
%C_s(p) \le 1-\h(p).
%$$
%\end{lemma}
%\begin{IEEEproof}
This is true because, a strongly-$p$-limited adversary strategy is to flip each symbol with probability $p$, independently. That is, the
adversary can always simulate a binary symmetric channel, whose capacity is $1-\h(p)$. 
%\end{IEEEproof}

%There is another definition of capacity that asks the probability of error to decay exponentially. Namely,  let,
%
%\begin{equation}
%C^{\rm exp}_s(p)    \equiv \inf_{\epsilon>0} \limsup_{n\to \infty} \frac{\log M^s_{\exp(-\epsilon n)}(n,p)}{n}.
%\end{equation}
%
%Again, it is clear that,
%\begin{equation}
%C^{\rm exp}_s(p)  \le C_s(p) \le 1-\h(p).
%\end{equation}
%
%Our main result is the following.
%
%\begin{theorem}\label{thm:strong}
%$C^{\rm exp}_s(p)$ is strictly less than $1-\h(p)$. Namely, there exists $p^{\ast} < \frac12$ such that
%$C^{\rm exp}_s(p)= 0$ for all $p \ge  p^{\ast}$.
%\end{theorem}

%It is easy to see that, the worst-case (over codewords) probability of error would not lead to 
%anything other than classical rate

\subsection{Practical limitations to the model and contributions} It is counterintuitive to assume that the  adversary,
being memoryless, cannot store the previously transmitted bits, or its own actions, however,
has access to the entire code and can do computations on them. But it should be noted that, the entire computation of the adversary 
is done offline, and in each transmission, it just performs according to  one of the two options.
Also note that, the adversary knows the time-instance of the transmission. That is, he knows that the $i$th transmission, among the
$n$ possible, is taking place. In that sense the adversary is not completely memoryless.
The main purpose of introducing this
model is to see how weak the adversary can be and still have its  capacity dictated  by the Plotkin bound.  

On the other hand, the concept of such memoryless adversary appears in principle before in literature.
In particular, general classes of restricted adversarial channels were considered in the literature of
 {\em arbitrarily varying channels} \cite{ahlswede1970capacity,csiszar1989capacity, csiszar1988capacity} or {\em oblivious channels} \cite{langberg2008oblivious}.
 From \cite{guruswami2013optimal} (see also,\cite{ahlswede1978arbitrary}), Thm.~C.1, it is evident that the capacity of weakly-$p$-limited
 adversary is $0$ for $p>\frac14$. It is also proved there that, if the adversary can keep a count of how many
 bits it has flipped (a log-space channel), then the same  fact holds for strongly limited adversaries as well.
 
 In  Sec.~\ref{sec:weak}, we present the above fact regarding weakly-limited adversary in a way 
 that is amenable to our definitions. We then attempt to extend this result to the case of strongly-limited 
 adversary: what we have forms the main contribution of this paper. In Sec.~\ref{sec:dist} we introduce the important notions of distance distribution of
  a code that proves useful in this context. In Sec.~\ref{sec:strong}, we show that the capacity of a strongly-$p$-limited
  adversary is strictly separated from the capacity of a BSC($p$). In particular we give an upper bound on $C_s(p)$ that is
  strictly below $1-\h(p)$ for all $p> \frac14$. Further discussions and concluding remarks
  are presented in Sec.~\ref{sec:skew}.

\section{Weakly-limited adversary}\label{sec:weak}
%We prove the following result.
%\begin{theorem}The capacity of weakly-limited memoryless adversary must follow,
%\begin{equation}
%%C(p) \le \min \{1-4p, 1-\h(p)\},
%C(p) \le 1-4p.
%\end{equation}
%%where $\h(x) = -x \log x -(1-x)\log(1-x)$ is the {\em binary entropy function}.
%\end{theorem}
%
%The theorem is proved using a series of lemmas.
%\begin{lemma}
%$$
%C(p) \le 1-\h(p).
%$$
%\end{lemma}
%\begin{IEEEproof}
%The adversary can flip each symbol with probability $p$, independently. That is, the
%adversary can always simulate a binary symmetric channel, whose capacity is $1-\h(p)$. 
%\end{IEEEproof}
In this small section, we establish the  following fact. % that will lead to the main questions of the paper.
\begin{theorem}\label{thm:weak}
$C_w(p) =0$ for $p \ge \frac14$.
\end{theorem}

To prove the theorem, the below lemma, known as the Plotkin bound, is used crucially.

\begin{lemma}[Plotkin Bound]\label{lem:plotkin}
Suppose, $\cC\subseteq \ff_2^n$ is the code and $|\cC|=M$. Randomly and uniformly (with replacement) choose two
codeword $\bfx_1,\bfx_2$ from $\cC$. Then,
\begin{equation}
\avg \dist(\bfx_1,\bfx_2) \le \frac{n}2,
\end{equation}
where $\dist(\cdot)$ is the Hamming distance.
\end{lemma}
\begin{IEEEproof}
Consider an $M \times n$ matrix with the codewords of $\cC$ as its rows. Suppose, $\lambda_i$ is the
number of $1$s in the $i$th column of the matrix,  $i=1,\dots , n$. Then, 
$$
\sum_{\bfc_1,\bfc_2 \in \cC} \dist(\bfc_1,\bfc_2) = 2\sum_{i=1}^n \lambda_i(M-\lambda_i) \le \frac{nM^2}{2}.
$$
Hence,
$
\avg \dist(\bfx_1,\bfx_2) \le \frac{n}2,
$
where, $\bfx_1,\bfx_2$ are two randomly and uniformly chosen codewords. 

\end{IEEEproof}

\begin{IEEEproof}[Proof of Theorem \ref{thm:weak}]
We show that there exists an adversary strategy that achieves the claim of the lemma.
In this vein, we use the same adversarial strategy that is used in \cite{guruswami2013optimal,dey2013upper}.
Suppose, $\cC\subset\ff_2^n$ is the code and $|\cC|=M$.  The adversary (channel) first choses
a codeword $\bfx = (x_1,x_2,\dots, x_n) \in \ff_2^n$ randomly and uniformly from $\cC$.
Now if $\bfc =(c_1,c_2,\dots,c_n)$ is the transmitted codeword, then,
\[
e_i\equiv  f_{\cC}^i(x_i) = 
\begin{cases}
    0,& \text{ when } x_i =  c_i\\
    1,  & \text{ with probability } \frac12 \text{ when } x_i \ne c_i\\
    0,  & \text{ with probability } \frac12 \text{ when } x_i \ne c_i.
\end{cases}
\]
Note that, if $\bfc$ is randomly and uniformly chosen from $\cC$, then
\begin{align*}
\avg \wt(\bfe) = \sum_{i=1}^n \Pr(e_i =1)& =\frac12\sum_{i=1}^n \Pr(x_i \ne c_i)\\
& = \frac12\avg \dist(\bfx,\bfc)  \le \frac{n}{4},
\end{align*}
where, $\bfe =(e_1,\dots,e_n)$. Hence, the adversary is weakly-$\frac14$-limited.
%
%And, hence,
%$$
%\Pr\Big(\wt(\bfe)< n\Big(\frac14+\epsilon\Big)\Big) > \frac{4\epsilon}{1+4\epsilon},
%$$
%from Markov inequality.

On the other hand, $\Pr(\bfx =\bfc) = \frac1M$. Suppose, $\bfy = \bfx+\bfe$. 
At the decoder,
let $\Pr(\bfy \mid \bfc')$, $\bfc' \in \cC$, denote the probability
that $\bfc'$ is transmitted and $\bfy$ is received.  
Clearly,
$$
\Pr(\bfy \mid \bfc) =\Pr(\bfy \mid \bfx).
$$
Hence, even with the maximum likelihood decoder will have a probability of error $\ge 1/2-\frac1M$.
Therefore, $C_w(p) =0$ for $p\ge 1/4$.
\end{IEEEproof}
%\begin{lemma}
%\begin{equation}
%M_\epsilon(n,p) \le 2 M_\epsilon(n-1,p).
%\end{equation}
%\end{lemma}
%\begin{IEEEproof}
%Suppose, $M_\epsilon(n,p) > 2 M_\epsilon(n-1,p)$. Let $\cC$ be the code that achieves the
%size $M_\epsilon(n,p)$. Let $\cC^1$ be the subcode of $\cC$ that contains all codewords with the last 
%symbol equal to $1$. That is $ \cC^1 = \{\bfx = (x_1,\dots, x_{n}) \in \cC:  x_n =1\}$. Suppose, $\cC^0 = \cC\setminus \cC^1$.
%Clearly, either $|\cC^1|$ or $|\cC^0|$ is greater than or equal to $|\cC|/2$. Without loss of generality, assume
%$|\cC^1|\ge |\cC|/2.$ But that means $|\cC^1| > M_\epsilon(n-1,p)$. As, the last bit of every vector in $\cC^1$ is equal (to $1$),
%we can puncture the last coordinate (i.e., get rid of this coordinate) to obtain a code of length $n$ and size $> M_\epsilon(n-1,p)$.
%Therefore, there exist a strategy in ${\rm Adv}(\cC^1,p)$.  
%

%\subsection*{Time sharing}
%The capacity is $\le 1- 4p$. This is true because the adversary may decide to do nothing in for the first $(1-4p)n$.

\section{Distance distribution}\label{sec:dist}
To extend Thm. \ref{thm:weak} to the case of strongly-limited adversary, we need to show an adversary strategy, 
that, with high probability, keep the number  of errors within $pn$. However, for the adversary strategy 
of Thm. \ref{thm:weak} to do this, we need the result of Lemma \ref{lem:plotkin} to be stronger, i.e., a high probability
statement.
%\begin{theorem}\label{thm:strong}
%\[
%C_s(p)
%\begin{cases}
%     = 0,& \text{ when } p \ge \frac14\\
%    \le 1-\h(p),  & \text{ when } p < \frac14 ,
%\end{cases}
%\]
%where,  $\h(x) = -x \log x -(1-x)\log(1-x)$ is the {\em binary entropy function}.
%\end{theorem}
Let us now introduce some notations that help us cast Lemma \ref{lem:plotkin} as a high-probability result.

The {\em distance distribution} of a code is defined in the following way. Suppose, $\cC \subseteq \ff_2^n$
be a code. Let, for $i=0,1,2,\dots, n$,
\begin{equation}
A_i = \frac1{|\cC|} |\{(\bfc_1,\bfc_2) \in \cC^2: \dist(\bfc_1,\bfc_2) = i \}|.
\end{equation}
As can be seen, $A_0 =1$.

The dual distance distribution of a code is defined to be, for $i =0,1,\dots, n$,
\begin{equation}
A^{\perp}_i = \frac1{|\cC|}\sum_{j=0}^n K_i(j) A_j,
\end{equation}
where $$K_i(j) = \sum_{k=0}^i (-1)^k \binom{j}{k}\binom{n-j}{i-k}$$
is the Krawtchouk polynomial.  Note that, $A^{\perp}_0 = 1$. It is known that
$A^{\perp}_i \ge 0$ for all $i$.
The {\em dual distance} $d^{\perp}$ of the code is defined to be the smallest  $i>0$ such that
$A^{\perp}_i$ nonzero.

\bigskip

\begin{lemma}[Pless power moments]\label{lem:pless}
For all  $r <d^{\perp}$,
\begin{equation}
\frac{1}{|\cC|} \sum_{i=0}^n (n/2- i)^r A_i = \frac{1}{2^n}\sum_{i=0}^n (n/2 -i)^r \binom{n}{i}.
\end{equation}
\end{lemma}
\begin{IEEEproof}
For a proof of the lemma, see \cite[p.~132]{MS1977}.
\end{IEEEproof}

\bigskip

\begin{lemma}\label{lem:conc}
Suppose, $\cC\subseteq \ff_2^n$ is the code  with dual distance greater than $2$, and $|\cC|=M$. Randomly and uniformly (with replacement) choose two
codeword $\bfx_1,\bfx_2$ from $\cC$. Then,
\begin{equation}
\Pr\Big(\dist(\bfx_1,\bfx_2)< n(1/2+\epsilon)\Big) > 1- \frac{1}{4n\epsilon^2} .
\end{equation}
\end{lemma}
\begin{IEEEproof}
From Lemma \ref{lem:pless}, for any $r < d^{\perp}$,
\begin{align*}
\Pr\Big(\dist(\bfx_1,\bfx_2)\ge n(1/2+\epsilon)\Big) &\le \frac{\avg(\dist(\bfx_1,\bfx_2) - n/2)^r }{n^r\epsilon^r}\\
&= \frac{\frac{1}{2^n}\sum_{i=0}^n (n/2 -i)^r \binom{n}{i}}{n^r\epsilon^r}. 
\end{align*} 
In particular, substituting $r=2$ we have,
\begin{align*}
\Pr\Big(\dist(\bfx_1,\bfx_2)\ge n(1/2+\epsilon)\Big) \le \frac{n/4}{n^2\epsilon^2}
& = \frac{1}{4n\epsilon^2}.
\end{align*}
\end{IEEEproof}

The implication of the above result is following.
For any code $\cC$ with dual distance greater than $2$, there exists a strongly-$p$-limited adversary strategy such
that, probability of error is at least $\frac12-\frac1{|\cC|}$ for all $p\ge \frac14$. The proof follows along the
lines of Thm.~\ref{thm:weak}.
However, this does not mean that the capacity of strongly-$p$-limited adversary becomes
$0$ for $p >\frac14$. There may exist a code with dual distance less than or equal to $2$ that
can reliably transfer information at a nonzero rate for $p >\frac14$. On the other hand, if the dual distance 
is that small, then the code must have a skewed or asymmetric distance distribution. In the next section, we
will (formally) see that this fact forces the capacity of the strongly limited adversary to be
strictly below that of binary-symmetric channel\footnote{It is known that the distribution of symbols (and even higher order strings) in the codebook needs to be {\em close}  to
the mutual information maximizing input distribution, such as uniform in BSC, for the code to achieve capacity
(see \cite{shamai1997empirical}). However, distance distribution is different than input distribution; and  we also want to quantify the
gap to capacity.}. 

\section{Strongly-limited adversary}\label{sec:strong}
The  main result of the paper concerns the capacity of strongly limited
adversary and is given in the following theorem.
\begin{theorem}\label{thm:strong}
\begin{equation}
C_s(p)\le \begin{cases}
 1-\h(p), \quad p \le \frac14\\
\h(1-3p+4p^2) -\h(p), \quad \frac14 < p \le \frac12.
\end{cases}
\end{equation}
\end{theorem}
To show this, we need to show the existence of an apt adversarial strategy.
\subsection{The adversary strategy}

The adversary uses the following strategy.

\begin{itemize}
\item $p \le \frac14$. The adversary just randomly and independently flips every bit with probability 
$p$.

\item $p > \frac14$. For the used code $\cC$, the adversary calculates   $L_\cC(p,n) = \sum_{w > 2pn} A_w,$
where $A_w$ is the distance distribution of the code. The following two cases may occur.
\begin{enumerate}
\item $\frac{L_\cC(p,n)}{|\cC|} = o(1).$ This case can be
 tested \footnote{Indeed, whenever we talk about a code, we mean a code-family, that is indexed by $n$, the length. In this case, the adversary knows this code family. There is a way to 
 bypass the $o(\cdot)$ notation, that we omit here for clarity.} 
if for any absolute constant $\epsilon$,  $\frac{L_\cC(p,n)}{|\cC|} <\epsilon$
for sufficiently large $n$. 
In this case, the adversary  first choses
a codeword $\bfx = (x_1,x_2,\dots, x_n) \in \ff_2^n$ randomly and uniformly from $\cC$.
Now if $\bfc =(c_1,c_2,\dots,c_n)$ is the transmitted codeword, then,  errors are introduced in the following way  
\[
e_i\equiv  f_{\cC}^i(x_i) = 
\begin{cases}
    0,& \text{ when } x_i =  c_i\\
    1,  & \text{ with Prob. } \frac12 \text{ when } x_i \ne c_i\\
    0,  & \text{ with Prob. } \frac12 \text{ when } x_i \ne c_i.
\end{cases}
\]
Let, $\bfe = (e_1,e_2,\dots,e_n)$. The received codeword is $\bfc+\bfe$.
\bigskip
\item $\frac{L_\cC(p,n)}{|\cC|} \ge c$ for some absolute constant $c$ for all $n$. In this case, 
the adversary just randomly and independently flips every bit with probability 
$p$.

\end{enumerate}
\end{itemize}

\subsection{Proof of Thm.~\ref{thm:strong}}
The following lemma will be useful in proving the theorem.
\begin{lemma}[Capacity of constrained input]\label{lem:const}
Let $R^{\ast}(p,\omega)$ denote the supremum of all achievable rates for a code (of length $n$) as $n \to \infty$ such 
that:
\begin{enumerate}
 \item Each codeword has Hamming weight at most $\omega n$, $\omega \le \frac12$.
 \item The average probability of error of using this code over BSC($p$) goes to $0$ as $n \to \infty$.
 \end{enumerate} 
Then $$
R^{\ast}(p,\omega) = \h(\omega\Asterisk p) -\h(p),  $$
where $\omega \Asterisk p = (1-\omega)p + \omega(1-p)$.
\end{lemma}
\begin{IEEEproof}[Sketch of proof]
To prove this lemma, we calculate the mutual information between the input and output of the BSC($p$),
when the input is i.i.d. Bernoulli($\omega$) random variables. It is  not difficult to show that, such random code
must contain almost as large a subset with weight of all codewords less than or equal to $\omega n$. The converse follows from 
an application of
Fano's inequality and noting that, asymptotically, $\log\binom{n}{\lambda n} \approx n \h(\lambda)$.
\end{IEEEproof}

\bigskip

\begin{IEEEproof}[Proof of Thm. \ref{thm:strong}]
If $p\le \frac14$ then the adversary just simulates the binary symmetric channel.
Below we consider the situation when $p > \frac14$. 

In what follows, we treat the two different scenarios for the adversary, based on the adversary strategy sketched above.
Let $\cC$ is the code that is used for transmission and $\{A_w\}$ is the
distance distribution of the code, as usual.

\bigskip

\noindent{\em Case 1:}
Let, $\bfx$ is the codeword adversary has initially chosen. Note that, if $\bfc$ is randomly and uniformly chosen from $\cC$, then,
the random variable $W= \dist(\bfc,\bfx)$ is distributed according to $\{A_w/|\cC|, w =0, \dots n\}$.

%Now, from \eqref{eq:case1}, 
%\begin{align*}
%\avg \exp(sW) &\le \varepsilon_n \exp\Big(\frac{sn}{2}(1+\epsilon)\Big) . %\\
%%\text{or,} \quad \sum_r \frac{s^r\avg(W-n/2)^r}{r!} & \le   \varepsilon_n \sum_r \frac{s^rn^r}{r!}.
%\end{align*}
%%That means there exists an integer $r$ such that,
%%$$
%%\avg{(W-n/2)^r} \le   \varepsilon_n  {n^r}.
%%$$
%

\bigskip

 Hence,
\begin{align*}
\Pr\Big(W > 2pn \Big) = o(1).
\end{align*}

%%$$
%%|\cC| \ge (1-2p)^{-n} \Big(\frac{2p}{1-2p}\Big)^{\frac{n(1+\epsilon)}{2}}.
%%$$
%%form Lemma \ref{lem:conc}, we have,
%%$$
%%\Pr\Big(\dist(\bfc,\bfx) \ge n/2(1+\epsilon)\Big) \le \frac{1}{n\epsilon^2}.
%%$$
Using Chernoff bound,
$$
\Pr\Big(\wt(\bfe)\ge  n (p+\epsilon) \Big| \dist(\bfc,\bfx) \le 2pn  \Big) \le  e^{-2 n\epsilon^2}.
$$
Hence, for any $\epsilon >0$,
$$
\Pr\Big(\wt(\bfe)<   n (p+\epsilon)   \Big) > 1- o(1),
$$
which imply that the adversary is strongly-$p$-limited.

Now, just following the arguments of Thm. \ref{thm:weak} we conclude that the code $\cC$ will
result in a probability of error at least $\frac12 -\frac1M$ with this adversary. 
Therefore,
If $C_s(p) >0$, then the next case must be satisfied for a code.  

\bigskip

\noindent{\em Case 2:}
%Suppose,  $\forall \frac12 <\alpha< 1$,
In this case,  there exists absolute constant $0<c<1$ such that,
\begin{equation}
\sum_{w>2p n} A_w \ge c|\cC|.
\end{equation}
For any codeword $\bfx \in \cC$, let $A_w^\bfx, w=0,\dots, n$ be the {\em local weight distribution}, i.e.,
the number of codewords that are at distance $w$ from $\bfx$.
Now as,
$$
\sum_{w>2p n} A_w = \frac{1}{|\cC|} \sum_{\bfx \in \cC} \Big(\sum_{w>2p n} A_w^\bfx \Big),
$$
it is clear that there must exist a codeword $\bfx$ such that
$$
\sum_{w>2p n} A_w^\bfx \ge c|\cC|.
$$
This ensures that, there are 
at least $c|\cC|$ codewords that belong within a Hamming ball of radius $n- 2pn = n(1-2p)$.
In particular, consider the ball of radius $n- 2pn$ centered at $\bar{\bfx}$, where $\bar{\bfx}$ is the
complement of $\bfc$ (all zeros are changed to ones, and vice versa). All the codewords of $\cC$ that are
distance more than $2pn$ away from $\bfx$ must belong to this ball; let us call the set of such codewords $\cB \subset \cC$. 
Clearly $|\cB| \ge c |\cC|$.

Consider the average probability of error, when $\cB$ is used to transmit a message over a BSC($p$).
Because, the Hamming space is translation invariant, the probability of error of such code is equal to
the probability of error of a code $\hat{\cB}$ that have the Hamming weight of each codeword bounded
by $n(1-2p)$. 
But from Lemma \ref{lem:const}, the maximum possible rate for which the probability of error of using $\cB$ in  BSC($p$) goes to
$0$ is $R^{\ast}(p, 1-2p)$.

However, if we randomly pick up a codeword from $\cC$, with probability at least $c>0,$ the codeword belong to $\cB$.
Hence $\frac1n \log |\cB|$ must be less than  $R^{\ast}(p, 1-2p)$, otherwise the average probability of error
for $\cC$ will be bounded away from $0$. Hence, the rate of $\cC$ is at most
$$
R^{\ast}(p, 1-2p) = \h(1-3p+4p^2) -\h(p).
$$

\end{IEEEproof}

\begin{figure}[t]
\begin{center}
\includegraphics[width=0.5\textwidth]{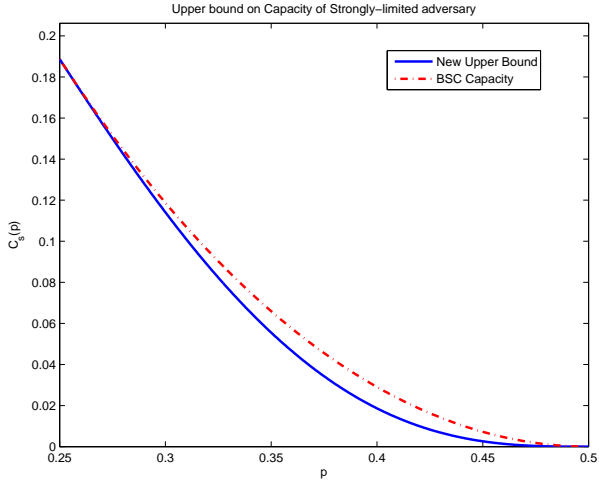}
\end{center}
\caption{The upper bound of Thm.~\ref{thm:strong} on the strongly-limited adversary.}
\label{fig:upper}
\end{figure}

The capacity of strongly-limited adversary is strictly bounded away from the capacity of BSC.
Indeed, $\h(1-3p+4p^2) <1$ for all $\frac14 < p < \frac12$. This is shown in Figure~\ref{fig:upper}.

\subsection{Erasure Channel}
The entire analysis of the above section can be extended for the case of a memoryless adversarial erasure channel,
where instead of corrupting a symbol, the adversary introduces an erasure.
Recently, an extension (that results in rather nontrivial observations) of the results of \cite{dey2013upper,haviv2011beating}
for the case of erasures have been performed in \cite{bassilycausal}.

We refrain from formally defining   a binary-input memoryless adversarial erasure channel; however, that can be done
easily along the lines of the introductory discussions of this paper. For the case of weakly-p-limited adversary the capacity is zero
for all $p \ge \frac12$. On the other hand, we note that, for strongly-p-limited adversarial erasure
channel the   capacity is upper bounded by $$(1-p)\h(p),$$ for all $p \ge \frac12$. The analysis is similar to 
that of this section, and uses the capacity of a constrained input erasure channel as a component of the proof (for example, see Eq.~7.15 of
\cite{cover2012elements}).
%For erasure channel we do the same analysis to find that the

\section{A code with skewed distance distribution}\label{sec:skew}
In conclusion we outline a possible route through which an improvement on 
the upper bound on $C_s(p)$ might be possible.

From the proof of Thm.~\ref{thm:strong} it is evident that a code $\cC$ that
has nonzero rate  can achieve a zero probability of error for the
strongly-$p$-limited adversary only if the distance distribution $\{A_w, w=0, \dots, n\}$ satisfies, for some absolute constant $c>0$,
 \begin{equation}\label{eq:skew}
\sum_{w>2p n} A_w \ge c|\cC|.
\end{equation}
From, Delsarte's theory of linear-programming bounds \cite{delsarte1973algebraic}, it is possible
to upper bound the maximum possible size of such code $\cC$. Indeed, 
this is given in the following theorem .

\bigskip

\begin{theorem}
Suppose, a code $\cC$ is such that its distance distribution  $\{A_w, w=0, \dots, n\}$ satisfies \eqref{eq:skew}
 for some $c>0$. Assume there exist a polynomial $f(x)$ of degree at most $n$  with,
 \begin{equation}
 f(x) = \sum_{k=0}^{n}f_k K_k(x),
 \end{equation}
and some $\beta>0$,  that  satisfy,
 \begin{enumerate}
 \item $f_0 =1$, $f_k \ge 0$ for $k =1\dots,n$;
 \item $f(j) \le c \beta$ for $j = 1,\dots, 2pn$ and  
 $f(j) \le -(1-c)\beta $ for $j = 2pn+1, \dots, n$.
 \end{enumerate}
 Then $$ |\cC| \le f(0) -c\beta.$$
 \label{thm:skew}
\end{theorem}
\begin{IEEEproof}
We note that, $A_i^\bot \ge 0$ for all $i =0, \dots, n$, a set of linear constraints on the 
distance distribution whose sum we want to maximize. Moreover we have the extra linear constraint of 
\eqref{eq:skew}.  We omit  the proof here, but if follows from standard arguments of linear programming bounds for codes.
\end{IEEEproof} 
If one could find a polynomial that satisfies the above conditions then 
that gives bounds on the capacity of strongly-$p$-limited adversary. Our current approach involves tweaking the
existing polynomials that bound error-correcting codes (i.e., the MRRW polynomials \cite{mceliece1977new})
to construct a polynomial that satisfy the criteria of Thm.~\ref{thm:skew}.
%\subsection*{Acknowledgement}
%This work was supported in part by NSF grant 1318093 and a grant from University of Minnesota.
%

\bibliographystyle{abbrv}
\bibliography{/Users/arya/Documents/BIB/aryabib}
\end{document}